\documentclass[amsmath,amssymb,aps,prb,preprint]{revtex4-2}

\usepackage[utf8]{inputenc}
\usepackage{graphicx,hyperref}
\usepackage{siunitx}
\usepackage{amsmath}
\usepackage{amssymb}
\usepackage[margin=2.7cm]{geometry}

\usepackage{ifthen}
\newboolean{revtex}
\makeatletter
\@ifclassloaded{revtex4-2}{
   \setboolean{revtex}{true}}{
   \setboolean{revtex}{false}
   }
\setboolean{revtex}{true}

\begin{document}
\ifthenelse{\boolean{revtex}}{
  \title{ de Haas van Alphen oscillations in hybridization-gap insulators 
          as a sudden change in the diamagnetic moment of Landau levels }
  \author{S.~R.~Julian}
  \affiliation{ 
      Department of Physics, University of Toronto, 
      60 St. George Street, Toronto, Ontario, Canada M5S 1A7 }
  \date{\today}
  }{
  \begin{center}
  {\bf \large  de Haas van Alphen oscillations in hybridization-gap insulators 
  as a sudden change in the diamagnetic moment of Landau levels \\ }
  {S.~R.~Julian \\ }
  {\it 
     Department of Physics, University of Toronto, \\
     60 St. George Street, Toronto, Ontario, Canada M5S 1A7 \\}
     (Dated: December 20, 2022) \\ 

  \vspace{5mm}  

  }

\ifthenelse{\boolean{revtex}}{\begin{abstract} }{
{\large Abstract} \end{center} }
This note revisits the semi-classical 
  theory of quantum oscillations in hybridization-gap 
  insulators, 
  and shows that 
  the physical origin of the oscillations, at $T=0$ K,  
   is a 
    sudden change in the diamagnetic 
     moment of each Landau level as it 
     crosses the hybridized region of the valence band. 
\ifthenelse{\boolean{revtex}}{ 
   \end{abstract} \maketitle \newpage
   }{\newpage}

\section{Introduction}
\label{sec-intro}
For more than 80 years 
  after the first observation of a quantum oscillatory 
  magnetization,  
  the de Haas-van Alphen (dHvA) effect 
   was regarded 
    as a signature of 
   a metal, 
because the theory was formulated 
  in terms of the 
    emptying of quantized 
     Landau levels as they 
      sequentially 
    cross the Fermi surface. 
In this picture, 
  there should be no oscillatory magnetization for 
  a filled or empty band, because there is no Fermi surface 
  for the Landau levels to cross. 
Thus, the observation 
   of dHvA oscillations in 
   the Kondo insulator SmB$_6$ 
   \cite{LiLi2014,Tan2015} 
   was greeted with suprise. 
Almost equally surprising, a possible, very simple, theory of these 
  quantum-oscillations-without-a-Fermi-surface 
  was soon found: 
Knolle and Cooper 
  \cite{Knolle2015} 
  showed that quantum oscillations emerge  
  in a straightforward calculation of the thermodynamic 
  potential of a narrow-gap insulator 
  with Landau-quantized states.  

The physical origin of these oscillations has,
  however,  
  been unclear, and 
  a source of some confusion. 
As  Ref.\ \cite{Ram2017} states, 
  in the absence of a Fermi surface  
  ``it is not clear what surface is being measured." 
The original Knolle and Cooper paper \cite{Knolle2015} 
  does not suggest a physical origin for the oscillations: 
  they emerge from the mathematics.  
Subsequent papers by these and other authors 
  \cite{Knolle2017,Pal2016,Pal2017,Panda22} 
  simply state that 
  the thermodynamic potential 
  oscillates as Landau levels sequentially cross the region where 
  the bands cross, 
  although Pal \cite{Pal2017} makes the enigmatic 
  remark that, 
  as $B$ changes, the Landau levels ``feel" the abrupt change 
  in the slope of $E(k)$ and ``manifest as quantum oscillations". 
Moreoever, some papers that offer competing theories, 
  e.g.\ \cite{Erten2016,Sodemann2018,Ghazaryan2021}, 
  state that the Knolle and Cooper dHvA oscillations 
  are due to magnetic breakdown, 
  in which the condition $\hbar\omega_c \gtrsim E_g^2/E_F$, 
  where $\omega_c$ is the quasiparticle cyclotron frequency, 
  $E_g$ is the energy gap and $E_F$ is the Fermi energy, 
  allows the quasiparticles to tunnel across the hybridization gap, 
  which would mean that these are, essentially, conventional quantum 
  oscillations.  

The purpose of this brief, somewhat pedagogical, note is to present a 
  simple semi-classical picture that shows that  
  the anomalous Knolle-Cooper dHvA oscillations in hybridzation gap insulators 
  are produced by the sudden change in the diamagnetic moment of the 
  Landau levels as they pass between regions of 
  $E(k)$ having different quasiparticle velocities.   
This is unrelated to magnetic breakdown, and indeed it shows that 
  this is a novel mechanism for the dHvA effect. 
This paper does not address the issue of whether this, as opposed to 
  one of several 
  competing theories (see Ref.\  
   \ifthenelse{\boolean{revtex}}{\onlinecite{Panda22}}{\cite{Panda22}}
  for a recent list), is 
  actually correct for the case of SmB$_6$. 

\section{Results and Discussion: Semiclassical calculation of the dHvA effect} 
\label{sec-fullySC} 

\begin{figure}
\begin{center}
\includegraphics[width=8.0cm]{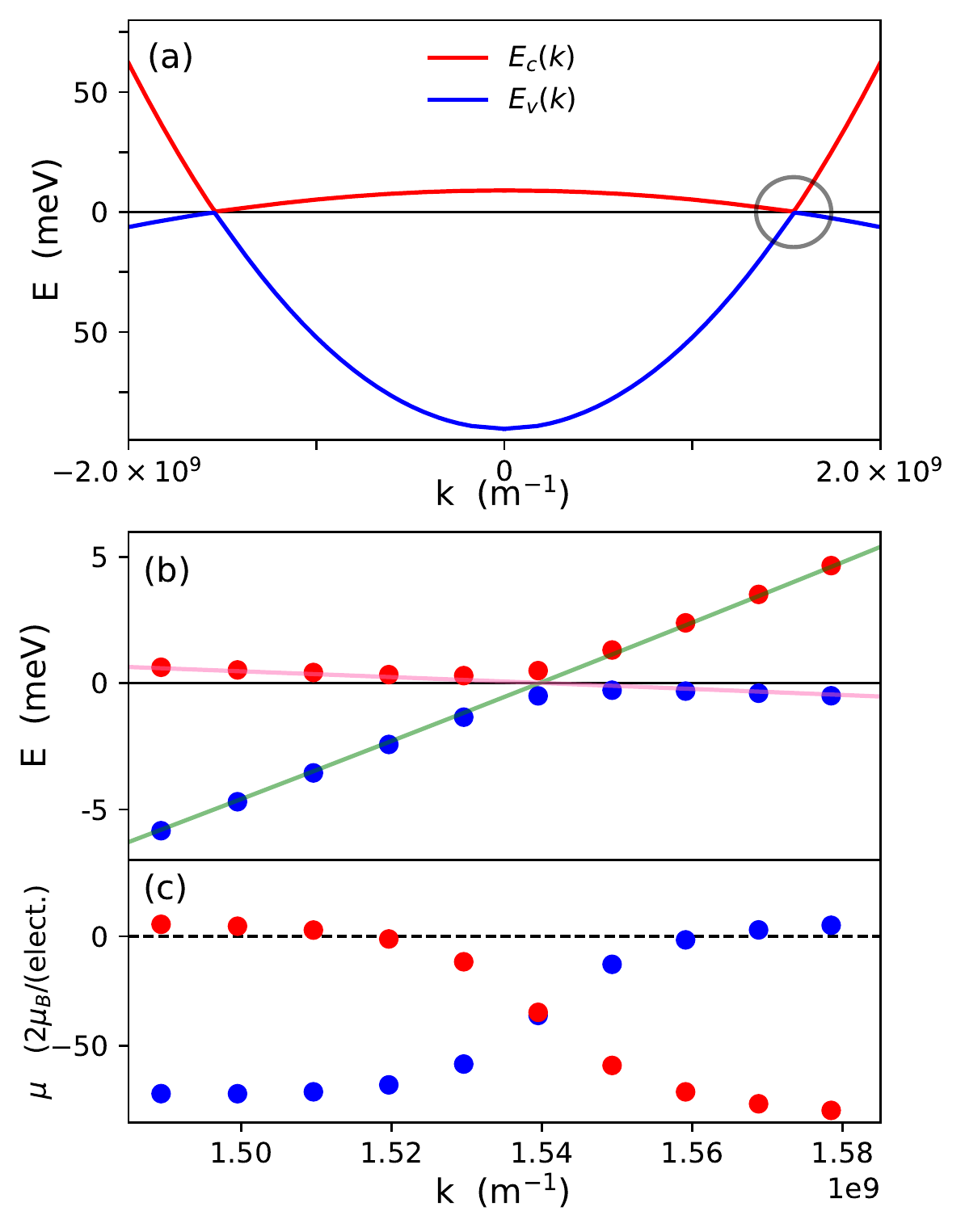}
\end{center}
\caption{(a) Band structure for a hybridization-gap insulator, 
        with the conduction band in red, and valence band in blue 
        (see \S \ref{sec-methods}, Methods, for details.)
  (b) A zoomed-in view of the circled region from (a), where the 
      bands cross and hybridize, with Landau 
      levels shown as blue and red points for the valence and 
      conduction band respectively. 
      The green and pink lines respectively are the unhybridized 
      electron and hole bands at zero field. 
  (c) Shows the diamagnetic moment per electron in the Landau 
      levels near the band crossing in the valence (blue) and 
      conduction (red) bands, in units of 
      double Bohr magnetons $\hbar e/m_e$. 
      The hybridization strength is moderate, so that the 
      band dispersion changes rather gradually compared to the 
      Landau level spacing. 
      At $T=0$ K the conduction-band Landau levels are unoccupied.
}
\label{fig-Fig1}
\end{figure}

The situation considered by Knolle and Cooper, 
  somewhat generalized in subsequent treatments 
        \cite{Knolle2015,Knolle2017,Pal2016,Pal2017,Panda22},  
  is as pictured in Fig.\ \ref{fig-Fig1}a. 
The essential feature is that two bands with 
  very different dispersions (in this case one electron-like and 
  one hole-like) 
  weakly hybridize where they cross, 
  and the Fermi energy $E_F$ lies in the gap.  
An applied magnetic field $B$ results 
  in the formation of Landau levels.  
As $B$ is increased, the 
  Landau levels, illustrated in Fig.\ \ref{fig-Fig1}b, 
  sequentially pass from the electron-like to the hole-like part 
  of the valence band, and vice-versa for the conduction band. 
Working, for simplicity, 
  at $T=0$ in two-dimensions with spinless fermions and no impurity scattering, 
  the grand canonical potential can be expressed as a sum 
  over all of the Landau levels of the valence band, 
\begin{eqnarray}
\Omega(\mu,T=0,B) = D \sum_{\ell}\left(E_{\ell,v} - E_F \right), 
  \label{eq-fully-qm}
\end{eqnarray}  
  where $D = (eB/2\pi \hbar)$ is the degeneracy of 
  each Landau level per unit area of sample, 
  and $E_{\ell,v}$ is the energy of the $\ell^{th}$ Landau level 
  in the valence band. 

In the conventional treatment of the dHvA effect in a metal 
  (see e.g.\ Ref.\  
  \ifthenelse{\boolean{revtex}}{\onlinecite{Shoenberg})}{\cite{Shoenberg})},
  the sum in Eq.\ \ref{eq-fully-qm} 
  is approximated in such a way that the oscillatory part, 
   $\tilde{\Omega}$, 
  which involves only states in the immediate vicinity of $E_F$, 
  can be extracted. 
Then it is straightforward to calculate the oscillations in 
  any thermodynamic quantity. 
For example, the dHvA effect measures the oscillatory magnetization 
\begin{eqnarray}
\tilde{M} = - \frac{\partial \tilde{\Omega}}{\partial B}. 
  \label{eq-tildeM}
\end{eqnarray}

One could approach the problem differently, however, 
  obtaining the total magnetization by 
  taking the derivative of $\Omega$, 
  rather than $\tilde{\Omega}$. 
Consider first a simple parabolic band of non-interacting electrons, 
  so that the dispersion relation is $E(k) = \hbar^2 k^2/2m$. 
Then, as in the case of electrons in free space, the 
  Landau levels are quantized in energy as 
\begin{eqnarray}
E_\ell = \hbar\omega_c(\ell+1/2), 
\end{eqnarray}
 where the cyclotron frequency is $\omega_c = eB/m$. 
Substituting this in Eq.\ \ref{eq-fully-qm}, and taking the derivative 
  with respect to $B$ using the non-obvious step of 
  applying the chain rule to separate the 
  derivative inside the sum from the derivative of the prefactor $D$, 
  gives 
\begin{subequations}
\begin{align}
M = &-D \sum_{\ell=0}^{\ell_{max}}\frac{e\hbar}{m}(\ell+1/2) \qquad \\ 
    & \qquad - \frac{D\hbar\omega_c}{B}\sum_{\ell=0}^{\ell_{max}} [(\ell+1/2) - 
                  X].
\end{align}
  \label{eq-diamagM}
\end{subequations} 
\noindent
The new variables are $X \equiv E_F/\hbar\omega_c$, 
  and 
  $\ell_{max}$, 
   which is the quantum number of the highest occupied 
  Landau level at a given field $B$, 
     which satisfies 
      $(\ell_{max}+1/2) \le E_F/\hbar\omega_c \le (\ell_{max} + 3/2)$.
$\ell_{max}$ changes by one every time a Landau level crosses $E_F$.  

\begin{figure}
\begin{center}
\includegraphics[width=8.0cm]{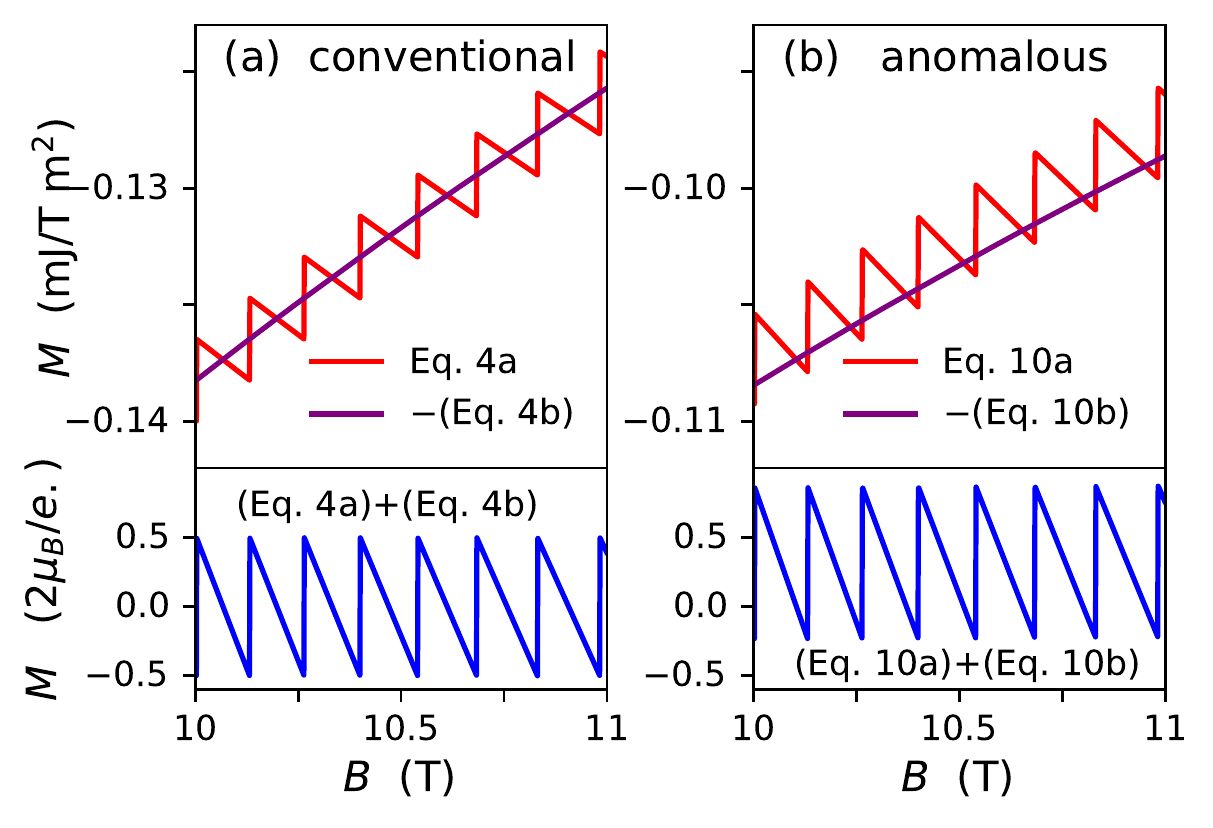}
\end{center}
\caption{
  (a), top panel, 
    plots Eq.\ \ref{eq-diamagM}a and the negative of \ref{eq-diamagM}b 
    for the conventional dHvA effect in an isolated metallic band of mass 
    $m = 1 m_e$, 
    showing that Eq.\ \ref{eq-diamagM}a
    contains the quantum oscillations, while \ref{eq-diamagM}b, 
    to a very good approximation, cancels the 
    quasi-linear background, leaving a total 
    magnetization (lower panel) that is dominated by the oscillations. 
    (See \S \ref{sec-methods}, Methods, for details.) 
 (b) shows the corresponding plot for Eq.\ \ref{eq-genM}, 
   the hybridization-gap insulator of Fig.\ \ref{fig-Fig1}, 
   with a very weak hybridization so that the band dispersion 
   changes very suddenly, compared to the Landau level spacing.
}
\label{fig-Fig2}
\end{figure}

In Fig.\ \ref{fig-Fig2}a the two terms of Eq.\ \ref{eq-diamagM} 
  are plotted separately (top), 
  and summed (bottom). 
It can be seen that, to a good approximation, 
  the first term is comprised of an oscillatory part superposed on 
  a large background, while the second term cancels (or very nearly cancels) 
  the large background of the first term. 
(The second term also has an oscillatory part, but 
  it is normally ignored, being 
  typically less than 1\% of the oscillatory part of the first term.) 
That is, the oscillations are almost entirely contained in the first term,
 which has a simple physical interpretation: 
  the 
  Landau diamagnetic dipole moment of an electron in the $\ell^{th}$ 
  Landau level is $\mu = -(e\hbar/m)(\ell+1/2)$ (see below). 
Thus the oscillatory magnetization comes from the 
  sum over the diamagnetic dipole moments of the electrons in 
  their Landau levels, and  
  in particular the sharp step in the `sawtooth' pattern of 
  $M$ is due to the loss of the Landau diamagnetic moment of 
  the electrons in the $\ell_{max}$ Landau level, 
  when the level suddenly empties as 
  it passes $E_F$ with increasing $B$. 

Shoenberg, in his book,  
  \cite{Shoenberg} 
  credits Brian Pippard with this insight, but 
  it must have been known to Landau and others. 
Pippard's treatment, 
  contained in 
  the proceedings of a summer school 
  that was held in Vancouver, Canada, in 1967 \cite{Pippard68},  
  is nevertheless useful. 
In a departure from the usual Landau gauge approach, 
  he uses the cylindrical gauge to construct 
  wave-functions, for the Landau quantized electrons, 
  that in real space are sharply 
  peaked (provided $\ell$ is not too small) at the classical 
  cyclotron radius, which encloses a real-space area $A_r$. 
Since the electrons circulate with period $T = 2\pi/\omega_c$, 
  their diamagnetic dipole moment is 
\begin{eqnarray}
\mu = -\frac{e\omega_c}{2\pi}A_r = -\frac{e^2B}{2\pi m}A_r. 
\end{eqnarray}
Pippard merely remarks that one can reformulate the dHvA effect 
  directly in terms of the diamagnetic magnetization. 

For an arbitrary band structure, 
  applying the semi-classical relation between the 
  real-space, ${\cal A}_{r,\ell}$, 
  and the $k$-space, ${\cal A}_{k,\ell}$,  
  areas of a Landau orbit \cite{Shoenberg,AandM}, 
\begin{eqnarray}
A_{r,\ell}(B) = \frac{\hbar^2}{e^2 B^2} {\cal A}_{k,\ell}(B) 
            = \frac{2\pi\hbar}{e B} (\ell + \gamma), 
  \label{eq-areas}
\end{eqnarray}
where $\gamma = 1/2$ for circularly-symmetric two-dimensional bands 
such as we use here \cite{Roth66}, 
  gives 
\begin{eqnarray}
\mu = - \frac{\hbar e}{ m_\ell^*(B)} (\ell + \gamma), 
  \label{eq-mu}
\end{eqnarray}
where 
\begin{eqnarray}
  m_\ell^*(B) \equiv \frac{\hbar^2}{2\pi}
               \left.\frac{\partial {\cal A}_k}{\partial E}\right|_{\ell,B}.
  \label{eq-mstar}
\end{eqnarray}

It is the factor $1/m_\ell^*$ in Eq.\ \ref{eq-mu} that is of interest. 
The sign of $m_\ell^*$ determines the direction in which 
  an electron circulates in 
  its Landau orbit, and thus determines the sign of the dipole moment; 
  meanwhile the magnitude of $m^*$, which in this case is really 
  a proxy for the speed of the electrons in the Landau orbit, determines 
  the magnitude of the moment: fast electrons (low $m_\ell^*$) 
  produce a strong diamagnetic moment, 
  slow electrons (high $m_\ell^*$) produce a weak diamagnetic moment. 
The diamagnetic moments of the Landau levels near the hybridization 
  gap are illustrated in Fig.\ \ref{fig-Fig1}c.
On the light electron-like part of the bands 
  the circulating electrons have a large negative diamagnetic moment, 
  but on the heavy hole-like part (where $m_\ell^*$ is negative) 
  they have a weak positive moment. 
The diamagnetic moment of the electrons in a 
  Landau level in the valence band thus undergoes a sudden change 
  when they cross the hybridized region, 
  which is located at the Fermi wave-vector of the 
  unhybridized bands. 

Returning to Eq.\ \ref{eq-fully-qm} and using 
\begin{eqnarray}
\frac{\partial E_{\ell}}{\partial B}  = 
   \frac{\partial E}{\partial {\cal A}_k} 
   \frac{\partial {\cal A}_k}{\partial B}  = 
   \frac{ e \hbar }{ m_\ell^*(B)}(\ell + \gamma), 
\end{eqnarray}
gives the generalized version of Eq.\ \ref{eq-diamagM}: 
\begin{subequations}
\begin{align}
M = &-D \sum_\ell^{occ.}\frac{\hbar e}{ m_\ell^*(B)} (\ell+\gamma) \qquad \\
    & \qquad 
     -D/B \sum_\ell^{occ.}(E_\ell - E_F).
\end{align}
  \label{eq-genM}
\end{subequations}
\noindent
In Fig.\ \ref{fig-Fig2}b the sum in 
  Eq.\ \ref{eq-genM} is evaluated for 
 the band structure of Fig.\ \ref{fig-Fig1}a. 
Despite the fact that the Landau levels do not cross the Fermi energy, 
  but rather remain in the valence band for all values of $\ell$, 
  it can be seen that the results for 
   the `anomalous' and `conventional' dHvA effects 
    are remarkably similar: 
the pattern is the same, the period of the oscillations is the same, and 
  again the oscillations are contained in the sum over the diamagnetic 
  moments of the electrons in the occupied Landau levels.
There are slight differences:  
  the magnitude of the quantum oscillations is slightly larger 
  in the anomalous case, 
  and the cancellation of background is not quite 
  as good. 

Eq.\ \ref{eq-genM}a, and Fig.\ \ref{fig-Fig1}c, 
  demonstrate the physical mechanism of 
  the Knolle-Cooper dHvA oscillations:  
  as each subsequent Landau level in the valence band 
  crosses the hybridized region at $k_F$, 
  its diamagnetic moment 
  undergoes a sudden change in sign and magnitude as the electrons suddenly 
  start circulating in the opposite sense at a different speed. 

The actual size of the step in $M$ depends on 
   the details of the band structure. 
In the case chosen here, 
  the velocity of the electrons in a Landau level reverses,  
  and slows by a factor of 10, as the Landau level passes from the 
  electron-like to the hole-like part of the valence band, 
  so the Landau diamagnetism goes from a negative value to 
  a 10$\times$ smaller, positive value, 
  so the step-change in $M$ is slightly larger 
  than in the conventional case, where 
  the diamagnetic moment merely disappears when the Landau level 
  empties as it crosses $E_F$. 
In the original Knolle and Cooper paper \cite{Knolle2015} the 
  hole band had infinite mass, so the diamagnetic moment 
  suddenly drops to zero at $k_F$ -- exactly as if the 
  Landau level had emptied. 

\begin{figure}
\ifthenelse{\boolean{revtex}}{}{ \begin{center} }
\includegraphics[width=8.0cm]{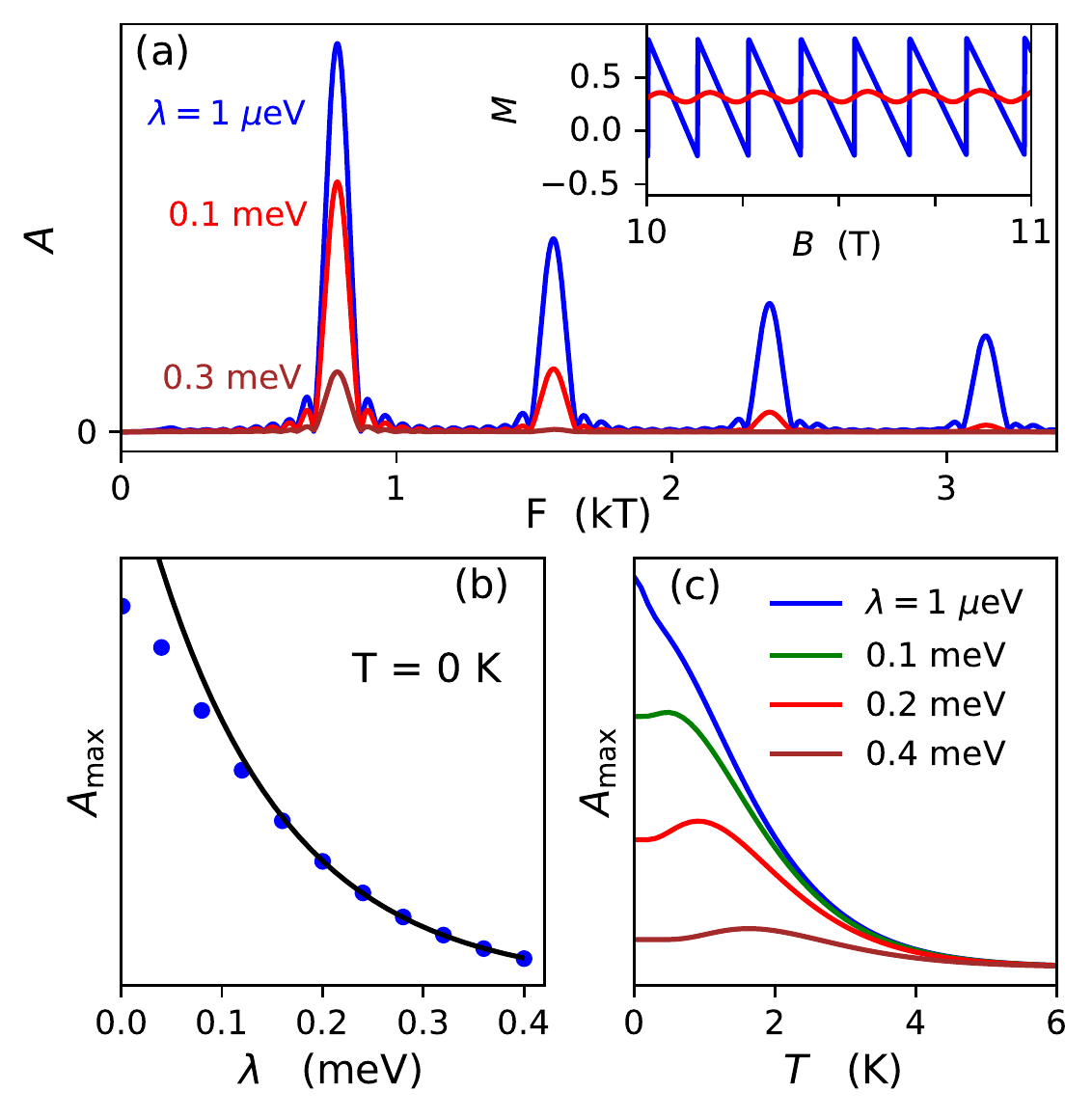}
\ifthenelse{\boolean{revtex}}{}{ \end{center} }
\caption{
 (a) shows the effect of changing hybridization strength. The inset 
  compares oscillations for $\lambda = 1$ $\mu$eV (blue line) with 
  $\lambda = 0.1$ meV (red line).  The main figure shows the Fourier 
  transform of $M$ vs $1/B$.  The frequency of the fundamental peak, 
   $F \sim 780$ T, 
  corrsponds via the Onsager relation, $F = \hbar {\cal A}_{k_F}/2\pi e$, 
  to the Fermi wave-vector of the unhybridized bands. 
  (See \S \ref{sec-methods}, Methods, for details.)
 (b) shows the amplitude of the fundamental peak vs. $\lambda$. 
  The black line is a decaying exponential fit to the points with 
    $\lambda > 0.15$ meV. 
 (c) shows the temperature dependence of the oscillations for 
  selected values of $\lambda$.
}
\label{fig-Fig3}
\end{figure}

In Figs.\ \ref{fig-Fig3}a and \ref{fig-Fig3}b 
  the effect of increasing the hybridization strength 
  $\lambda$ is shown.
There is an exponential fall in dHvA amplitude with increasing $\lambda$ 
  as long as $\lambda$ is not too small. 
This was noted by Knolle and Cooper, but 
  some authors \cite{Erten2016,Sodemann2018,Ghazaryan2021} mistakenly 
  attributed this 
  to magnetic breakdown, because it has a similar dependence on the 
  energy gap $E_g$. 
But in fact, 
  as noted in Ref.\  
  \ifthenelse{\boolean{revtex}}{\onlinecite{Pal2017}}{\cite{Pal2017}},
  the physical effect (recall that the conduction band is 
  unoccupied in this calculation) has to do with the {\em width} of the 
  hybridization-crossover region, compared with the 
  Landau level spacing.  
If one Landau level crossing the hybridized region 
  changes its diamagnetic moment before the next level 
  enters, there is a large oscillatory effect. 
If, on the other hand, several Landau levels are crossing this region 
  at the same time (the large $\lambda$ case), as pictured in 
  Fig.\ \ref{fig-Fig1}c, then the oscillations are small. 
It is of course a feature of degenerate perturbation theory 
  that the larger the gap, 
  the wider the region in $k$-space over which the bands change 
  their slope. 

Finally, in Fig.\ \ref{fig-Fig2}c the temperature dependence of the 
  peak in the amplitude spectrum for 
  several values of $\lambda$ is shown. 
Similar results were found by Knolle and Cooper, and discussed by them. 
The key point is that when $k_B T \gg E_g$ the hybridization gap no longer 
  matters, and conventional Lifshitz-Kosevich temperature dependence is found. 

\section{Conclusions}
\label{sec-concl}

By showing that the quantum oscillatory magnetization is 
  contained in the sum over the diamagnetic moments of the 
  electrons in occupied Landau levels, 
  it follows that at $T = 0$ K 
  the oscillations in 
  the anomalous dHvA effect arise,  
  not because the highest Landau level empties as in the 
  conventional dHvA effect, but rather because its 
  diamagnetic moment changes suddenly when the slope of the 
  valence band changes in the hybridization region. 
This gives a simple interpretation of the quantum 
  oscillatory magnetism found by Knolle and Cooper, and 
  reinforces that this is a novel mechanism for quantum oscillations. 

\section{Methods}
\label{sec-methods}

In Fig.\ \ref{fig-Fig1}a 
  two bands, with $E_1(k) = \hbar^2 k^2/2m_1$ and 
  $E_2(k) = W + \hbar^2 k^2/2m_2$, where $m_1 = 1m_e$, $m_2 = - 10 m_e$  
  and $W = 0.1$ eV, 
  are hybridized with a coupling of strength 
  $\lambda$. 
(The negative mass for $m_2$ is needed for consistency of notation, 
  particularly with Eq.\ \ref{eq-mstar}.) 
The resulting bands are 
\begin{eqnarray}
E_i(k) &=& 
    E_{av}(k) \pm \sqrt{ \Delta E(k)^2 + \lambda^2},
\end{eqnarray}
where $E_{av}(k) = 0.5(E_1(k) + E_2(k))$, 
      $\Delta E = 0.5(E_1(k) - E_2(k))$, 
  and  $E_i(k)$ is $E_c(k)$($E_v(k)$) for the $+$($-$) solution. 
Since the band structure is circularly symmetric,  
  the wave-vector corresponding to a given Landau level $\ell$ 
  at a field $B$ can be obtained from 
$\pi k_\ell(B)^2 = {\cal A}_{k,\ell}(B) = 2\pi e B(\ell+\gamma)/\hbar$, 
  and then $m_\ell^*(B)$ for $i = v,c$ is obtained from Eq.\ \ref{eq-mstar}.
In Figs.\ \ref{fig-Fig1}b and \ref{fig-Fig1}c the Landau levels are 
  evaluated at $B=10$ T. 

For Figs.\ \ref{fig-Fig2} and \ref{fig-Fig3},   
  Eqs.\ \ref{eq-mstar} and \ref{eq-genM} were numerically evaluated.  
A spherical Brillouin zone boundary was used, with a gradual logic-function  
  cutoff to avoid spurious zone-boundary dHvA oscillations. 
Care must be taken with this spherical zone boundary when 
  taking derivatives. 

For Fig.\ \ref{fig-Fig2}a, the hole-like band is absent, leaving 
  only a metallic band, with $m = 1m_e$. $E_F$ is the same as 
  for the anomalous dHvA case, so $k_F$ occurs at the crossing-point 
  of the unhybridized bands in Fig.\ \ref{fig-Fig1}b.
In Fig.\ \ref{fig-Fig2}b, lower panel, the same normalization was 
 used as in (a). That is, only electrons in the electron-like 
 part of the valence band are counted, 
 leaving out the electrons in the Landau levels of 
 the hole-like part, 
 which are of course also occupied. 
This makes comparison with Fig.\ \ref{fig-Fig2}a meaningful.

To get the temperature dependence for Fig.\ \ref{fig-Fig3}c the 
  usual expression for $\Omega$ was used: 
\begin{eqnarray}
\Omega = - D k_BT \sum_\ell 
             {\rm \ln}\left( 1 + {\rm e}^{(E_F - E_\ell)/k_BT}\right), 
\end{eqnarray}
which leads to the generalization of Eq.\ \ref{eq-genM}: 
\begin{subequations}
\begin{align}
M = &-D \sum_\ell 
         f(E_\ell,T)\frac{\hbar e}{ m_\ell^*(B)} (\ell+\gamma) \qquad \\
    & \qquad
    +\frac{k_B T D}{B} \sum_\ell{\rm \ln}\left( 1 + {\rm e}^{(E_F - E_\ell)/k_BT}\right), 
\end{align}
where $f(E_\ell,T)$ is the Fermi-Dirac distribution function, 
 and the sum  now includes both the conduction and the valence 
 bands. 
\end{subequations}

\section{Acknowledgements}

This research was funded by NSERC (RGPIN-2019-06446).

\ifthenelse{\boolean{revtex}}{
   \bibliographystyle{apsrev4-2}   
   \bibliography{bibby.bib}        

\begin{thebibliography}{15}%
\makeatletter
\providecommand \@ifxundefined [1]{%
 \@ifx{#1\undefined}
}%
\providecommand \@ifnum [1]{%
 \ifnum #1\expandafter \@firstoftwo
 \else \expandafter \@secondoftwo
 \fi
}%
\providecommand \@ifx [1]{%
 \ifx #1\expandafter \@firstoftwo
 \else \expandafter \@secondoftwo
 \fi
}%
\providecommand \natexlab [1]{#1}%
\providecommand \enquote  [1]{``#1''}%
\providecommand \bibnamefont  [1]{#1}%
\providecommand \bibfnamefont [1]{#1}%
\providecommand \citenamefont [1]{#1}%
\providecommand \href@noop [0]{\@secondoftwo}%
\providecommand \href [0]{\begingroup \@sanitize@url \@href}%
\providecommand \@href[1]{\@@startlink{#1}\@@href}%
\providecommand \@@href[1]{\endgroup#1\@@endlink}%
\providecommand \@sanitize@url [0]{\catcode `\\12\catcode `\$12\catcode
  `\&12\catcode `\#12\catcode `\^12\catcode `\_12\catcode `\%12\relax}%
\providecommand \@@startlink[1]{}%
\providecommand \@@endlink[0]{}%
\providecommand \url  [0]{\begingroup\@sanitize@url \@url }%
\providecommand \@url [1]{\endgroup\@href {#1}{\urlprefix }}%
\providecommand \urlprefix  [0]{URL }%
\providecommand \Eprint [0]{\href }%
\providecommand \doibase [0]{https://doi.org/}%
\providecommand \selectlanguage [0]{\@gobble}%
\providecommand \bibinfo  [0]{\@secondoftwo}%
\providecommand \bibfield  [0]{\@secondoftwo}%
\providecommand \translation [1]{[#1]}%
\providecommand \BibitemOpen [0]{}%
\providecommand \bibitemStop [0]{}%
\providecommand \bibitemNoStop [0]{.\EOS\space}%
\providecommand \EOS [0]{\spacefactor3000\relax}%
\providecommand \BibitemShut  [1]{\csname bibitem#1\endcsname}%
\let\auto@bib@innerbib\@empty
\bibitem [{\citenamefont {Li}\ \emph {et~al.}(2014)\citenamefont {Li},
  \citenamefont {Xiang}, \citenamefont {Yu}, \citenamefont {Asaba},
  \citenamefont {Lawson}, \citenamefont {Cai}, \citenamefont {Tinsman},
  \citenamefont {Berkley}, \citenamefont {Wolgast}, \citenamefont {Eo},
  \citenamefont {Kim}, \citenamefont {Kurdak}, \citenamefont {Allen},
  \citenamefont {Sun}, \citenamefont {Chen}, \citenamefont {Wang},
  \citenamefont {Fisk},\ and\ \citenamefont {Li}}]{LiLi2014}%
  \BibitemOpen
  \bibfield  {author} {\bibinfo {author} {\bibfnamefont {G.}~\bibnamefont
  {Li}}, \bibinfo {author} {\bibfnamefont {Z.}~\bibnamefont {Xiang}}, \bibinfo
  {author} {\bibfnamefont {F.}~\bibnamefont {Yu}}, \bibinfo {author}
  {\bibfnamefont {T.}~\bibnamefont {Asaba}}, \bibinfo {author} {\bibfnamefont
  {B.}~\bibnamefont {Lawson}}, \bibinfo {author} {\bibfnamefont
  {P.}~\bibnamefont {Cai}}, \bibinfo {author} {\bibfnamefont {C.}~\bibnamefont
  {Tinsman}}, \bibinfo {author} {\bibfnamefont {A.}~\bibnamefont {Berkley}},
  \bibinfo {author} {\bibfnamefont {S.}~\bibnamefont {Wolgast}}, \bibinfo
  {author} {\bibfnamefont {Y.~S.}\ \bibnamefont {Eo}}, \bibinfo {author}
  {\bibfnamefont {D.-J.}\ \bibnamefont {Kim}}, \bibinfo {author} {\bibfnamefont
  {C.}~\bibnamefont {Kurdak}}, \bibinfo {author} {\bibfnamefont {J.~W.}\
  \bibnamefont {Allen}}, \bibinfo {author} {\bibfnamefont {K.}~\bibnamefont
  {Sun}}, \bibinfo {author} {\bibfnamefont {X.~H.}\ \bibnamefont {Chen}},
  \bibinfo {author} {\bibfnamefont {Y.~Y.}\ \bibnamefont {Wang}}, \bibinfo
  {author} {\bibfnamefont {Z.}~\bibnamefont {Fisk}},\ and\ \bibinfo {author}
  {\bibfnamefont {L.}~\bibnamefont {Li}},\ }\href
  {https://doi.org/10.1126/science.1250366} {\bibfield  {journal} {\bibinfo
  {journal} {Science}\ }\textbf {\bibinfo {volume} {346}},\ \bibinfo {pages}
  {1208} (\bibinfo {year} {2014})}\BibitemShut {NoStop}%
\bibitem [{\citenamefont {Tan}\ \emph {et~al.}(2015)\citenamefont {Tan},
  \citenamefont {Hsu}, \citenamefont {Zeng}, \citenamefont {Ciomaga~Hatnean},
  \citenamefont {Harrison}, \citenamefont {Zhu}, \citenamefont {Hartstein},
  \citenamefont {Kiourlappou}, \citenamefont {Srivastava}, \citenamefont
  {Johannes}, \citenamefont {Murphy}, \citenamefont {Park}, \citenamefont
  {Balicas}, \citenamefont {Lonzarich}, \citenamefont {Balakrishnan},\ and\
  \citenamefont {Sebastian}}]{Tan2015}%
  \BibitemOpen
  \bibfield  {author} {\bibinfo {author} {\bibfnamefont {B.~S.}\ \bibnamefont
  {Tan}}, \bibinfo {author} {\bibfnamefont {Y.-T.}\ \bibnamefont {Hsu}},
  \bibinfo {author} {\bibfnamefont {B.}~\bibnamefont {Zeng}}, \bibinfo {author}
  {\bibfnamefont {M.}~\bibnamefont {Ciomaga~Hatnean}}, \bibinfo {author}
  {\bibfnamefont {N.}~\bibnamefont {Harrison}}, \bibinfo {author}
  {\bibfnamefont {Z.}~\bibnamefont {Zhu}}, \bibinfo {author} {\bibfnamefont
  {M.}~\bibnamefont {Hartstein}}, \bibinfo {author} {\bibfnamefont
  {M.}~\bibnamefont {Kiourlappou}}, \bibinfo {author} {\bibfnamefont
  {A.}~\bibnamefont {Srivastava}}, \bibinfo {author} {\bibfnamefont {M.~D.}\
  \bibnamefont {Johannes}}, \bibinfo {author} {\bibfnamefont {T.~P.}\
  \bibnamefont {Murphy}}, \bibinfo {author} {\bibfnamefont {J.-H.}\
  \bibnamefont {Park}}, \bibinfo {author} {\bibfnamefont {L.}~\bibnamefont
  {Balicas}}, \bibinfo {author} {\bibfnamefont {G.~G.}\ \bibnamefont
  {Lonzarich}}, \bibinfo {author} {\bibfnamefont {G.}~\bibnamefont
  {Balakrishnan}},\ and\ \bibinfo {author} {\bibfnamefont {S.~E.}\ \bibnamefont
  {Sebastian}},\ }\href {https://doi.org/10.1126/science.aaa7974} {\bibfield
  {journal} {\bibinfo  {journal} {Science}\ }\textbf {\bibinfo {volume}
  {349}},\ \bibinfo {pages} {287} (\bibinfo {year} {2015})}\BibitemShut
  {NoStop}%
\bibitem [{\citenamefont {Knolle}\ and\ \citenamefont
  {Cooper}(2015)}]{Knolle2015}%
  \BibitemOpen
  \bibfield  {author} {\bibinfo {author} {\bibfnamefont {J.}~\bibnamefont
  {Knolle}}\ and\ \bibinfo {author} {\bibfnamefont {N.~R.}\ \bibnamefont
  {Cooper}},\ }\href {https://doi.org/10.1103/PhysRevLett.115.146401}
  {\bibfield  {journal} {\bibinfo  {journal} {Phys.\ Rev.\ Lett.}\ }\textbf
  {\bibinfo {volume} {115}},\ \bibinfo {pages} {146401} (\bibinfo {year}
  {2015})}\BibitemShut {NoStop}%
\bibitem [{\citenamefont {Ram}\ and\ \citenamefont {Kumar}(2017)}]{Ram2017}%
  \BibitemOpen
  \bibfield  {author} {\bibinfo {author} {\bibfnamefont {P.}~\bibnamefont
  {Ram}}\ and\ \bibinfo {author} {\bibfnamefont {B.}~\bibnamefont {Kumar}},\
  }\href {https://doi.org/10.1103/PhysRevB.96.075115} {\bibfield  {journal}
  {\bibinfo  {journal} {Phys.\ Rev.\ B}\ }\textbf {\bibinfo {volume} {96}},\
  \bibinfo {pages} {075115} (\bibinfo {year} {2017})}\BibitemShut {NoStop}%
\bibitem [{\citenamefont {Knolle}\ and\ \citenamefont
  {Cooper}(2017)}]{Knolle2017}%
  \BibitemOpen
  \bibfield  {author} {\bibinfo {author} {\bibfnamefont {J.}~\bibnamefont
  {Knolle}}\ and\ \bibinfo {author} {\bibfnamefont {N.~R.}\ \bibnamefont
  {Cooper}},\ }\href {https://doi.org/10.1103/PhysRevLett.118.176801}
  {\bibfield  {journal} {\bibinfo  {journal} {Phys.\ Rev.\ Lett.}\ }\textbf
  {\bibinfo {volume} {118}},\ \bibinfo {pages} {176801} (\bibinfo {year}
  {2017})}\BibitemShut {NoStop}%
\bibitem [{\citenamefont {Pal}\ \emph {et~al.}(2016)\citenamefont {Pal},
  \citenamefont {Pi\'echon}, \citenamefont {Fuchs}, \citenamefont {Goerbig},\
  and\ \citenamefont {Montambaux}}]{Pal2016}%
  \BibitemOpen
  \bibfield  {author} {\bibinfo {author} {\bibfnamefont {H.~K.}\ \bibnamefont
  {Pal}}, \bibinfo {author} {\bibfnamefont {F.}~\bibnamefont {Pi\'echon}},
  \bibinfo {author} {\bibfnamefont {J.-N.}\ \bibnamefont {Fuchs}}, \bibinfo
  {author} {\bibfnamefont {M.}~\bibnamefont {Goerbig}},\ and\ \bibinfo {author}
  {\bibfnamefont {G.}~\bibnamefont {Montambaux}},\ }\href
  {https://doi.org/10.1103/PhysRevB.94.125140} {\bibfield  {journal} {\bibinfo
  {journal} {Phys.\ Rev.\ B}\ }\textbf {\bibinfo {volume} {94}},\ \bibinfo
  {pages} {125140} (\bibinfo {year} {2016})}\BibitemShut {NoStop}%
\bibitem [{\citenamefont {Pal}(2017)}]{Pal2017}%
  \BibitemOpen
  \bibfield  {author} {\bibinfo {author} {\bibfnamefont {H.~K.}\ \bibnamefont
  {Pal}},\ }\href {https://doi.org/10.1103/PhysRevB.95.085111} {\bibfield
  {journal} {\bibinfo  {journal} {Phys.\ Rev.\ B}\ }\textbf {\bibinfo {volume}
  {95}},\ \bibinfo {pages} {085111} (\bibinfo {year} {2017})}\BibitemShut
  {NoStop}%
\bibitem [{\citenamefont {Panda}\ \emph {et~al.}(2022)\citenamefont {Panda},
  \citenamefont {Banerjee},\ and\ \citenamefont {Randeria}}]{Panda22}%
  \BibitemOpen
  \bibfield  {author} {\bibinfo {author} {\bibfnamefont {A.}~\bibnamefont
  {Panda}}, \bibinfo {author} {\bibfnamefont {S.}~\bibnamefont {Banerjee}},\
  and\ \bibinfo {author} {\bibfnamefont {M.}~\bibnamefont {Randeria}},\ }\href
  {https://doi.org/10.1073/pnas.2208373119} {\bibfield  {journal} {\bibinfo
  {journal} {Proc.\ Natl.\ Acad.\ Sci.\ USA}\ }\textbf {\bibinfo {volume}
  {119}},\ \bibinfo {pages} {e2208373119} (\bibinfo {year} {2022})}\BibitemShut
  {NoStop}%
\bibitem [{\citenamefont {Erten}\ \emph {et~al.}(2016)\citenamefont {Erten},
  \citenamefont {Ghaemi},\ and\ \citenamefont {Coleman}}]{Erten2016}%
  \BibitemOpen
  \bibfield  {author} {\bibinfo {author} {\bibfnamefont {O.}~\bibnamefont
  {Erten}}, \bibinfo {author} {\bibfnamefont {P.}~\bibnamefont {Ghaemi}},\ and\
  \bibinfo {author} {\bibfnamefont {P.}~\bibnamefont {Coleman}},\ }\href
  {https://doi.org/10.1073/pnas.2208373119} {\bibfield  {journal} {\bibinfo
  {journal} {Phys.\ Rev.\ Lett.}\ }\textbf {\bibinfo {volume} {116}},\ \bibinfo
  {pages} {046403} (\bibinfo {year} {2016})}\BibitemShut {NoStop}%
\bibitem [{\citenamefont {Sodemann}\ \emph {et~al.}(2018)\citenamefont
  {Sodemann}, \citenamefont {Chowdhury},\ and\ \citenamefont
  {Senthil}}]{Sodemann2018}%
  \BibitemOpen
  \bibfield  {author} {\bibinfo {author} {\bibfnamefont {I.}~\bibnamefont
  {Sodemann}}, \bibinfo {author} {\bibfnamefont {D.}~\bibnamefont
  {Chowdhury}},\ and\ \bibinfo {author} {\bibfnamefont {T.}~\bibnamefont
  {Senthil}},\ }\href@noop {} {\bibfield  {journal} {\bibinfo  {journal}
  {Phys.\ Rev.\ B}\ }\textbf {\bibinfo {volume} {97}},\ \bibinfo {pages}
  {045152} (\bibinfo {year} {2018})}\BibitemShut {NoStop}%
\bibitem [{\citenamefont {Ghazaryan}\ \emph {et~al.}(2021)\citenamefont
  {Ghazaryan}, \citenamefont {Emilian}, \citenamefont {Erten},\ and\
  \citenamefont {Ghaemi}}]{Ghazaryan2021}%
  \BibitemOpen
  \bibfield  {author} {\bibinfo {author} {\bibfnamefont {A.}~\bibnamefont
  {Ghazaryan}}, \bibinfo {author} {\bibfnamefont {M.~N.}\ \bibnamefont
  {Emilian}}, \bibinfo {author} {\bibfnamefont {O.}~\bibnamefont {Erten}},\
  and\ \bibinfo {author} {\bibfnamefont {P.}~\bibnamefont {Ghaemi}},\ }\href
  {https://doi.org/10.1088/1367-2630/ac4124} {\bibfield  {journal} {\bibinfo
  {journal} {New J. Phys.}\ }\textbf {\bibinfo {volume} {23}},\ \bibinfo
  {pages} {123042} (\bibinfo {year} {2021})}\BibitemShut {NoStop}%
\bibitem [{\citenamefont {Shoenberg}(1984)}]{Shoenberg}%
  \BibitemOpen
  \bibfield  {author} {\bibinfo {author} {\bibfnamefont {D.}~\bibnamefont
  {Shoenberg}},\ }\href@noop {} {\emph {\bibinfo {title} {Magnetic oscillations
  in metals}}}\ (\bibinfo  {publisher} {Cambridge University Press},\ \bibinfo
  {year} {1984})\BibitemShut {NoStop}%
\bibitem [{\citenamefont {Pippard}(1968)}]{Pippard68}%
  \BibitemOpen
  \bibfield  {author} {\bibinfo {author} {\bibfnamefont {A.}~\bibnamefont
  {Pippard}},\ }\href@noop {} {\emph {\bibinfo {title} {Solid State Physics,
  vol.\ 1, Electrons in Metals}}},\ edited by\ \bibinfo {editor} {\bibfnamefont
  {J.~F.}\ \bibnamefont {Cochran}}\ and\ \bibinfo {editor} {\bibfnamefont
  {R.~R.}\ \bibnamefont {Haering}}\ (\bibinfo  {publisher} {Gordon and
  Breach},\ \bibinfo {year} {1968})\BibitemShut {NoStop}%
\bibitem [{\citenamefont {Ashcroft}\ and\ \citenamefont
  {Mermin}(1976)}]{AandM}%
  \BibitemOpen
  \bibfield  {author} {\bibinfo {author} {\bibfnamefont {N.~W.}\ \bibnamefont
  {Ashcroft}}\ and\ \bibinfo {author} {\bibfnamefont {N.~D.}\ \bibnamefont
  {Mermin}},\ }\bibinfo {title} {Solid state physics}\ (\bibinfo  {publisher}
  {Hold, Rinehart and Winston},\ \bibinfo {year} {1976})\ Chap.\ \bibinfo
  {chapter} {12 and 14}\BibitemShut {NoStop}%
\bibitem [{\citenamefont {Roth}(1966)}]{Roth66}%
  \BibitemOpen
  \bibfield  {author} {\bibinfo {author} {\bibfnamefont {L.}~\bibnamefont
  {Roth}},\ }\href {https://doi.org/10.1103/PhysRev.145.434} {\bibfield
  {journal} {\bibinfo  {journal} {Phys.\ Rev.\ B}\ }\textbf {\bibinfo {volume}
  {145}},\ \bibinfo {pages} {434} (\bibinfo {year} {1966})}\BibitemShut
  {NoStop}%
\end{thebibliography}%
   }{
}  

\end{document}